\documentclass[english]{article}
\usepackage[T1]{fontenc}
\usepackage[latin9]{inputenc}
\usepackage{geometry}
\usepackage{color}
\usepackage{array}
\usepackage{amstext}
\usepackage{graphicx}
\usepackage{epsfig}
\usepackage{epstopdf}
\makeatletter

\providecommand{\tabularnewline}{\\}

\@ifundefined{definecolor}
 {\usepackage{color}}{}
\makeatother

\makeatother

\usepackage{babel}

\begin{document}

\title{Controlled bidirectional remote state preparation in noisy environment:
A generalized view}

\author{Vishal Sharma$^{a}$, Chitra Shukla$^{b},$ Subhashish Banerjee$^{a}$,
Anirban Pathak$^{b,}$%
\thanks{anirban.pathak@jiit.ac.in%
} }

\maketitle
\begin{center}
$^{a}$Indian Institute of Technology Jodhpur, Rajasthan-342011, India 
\par\end{center}

\begin{center}
$^{b}$Jaypee Institute of Information Technology, A 10, Sector-62,
Noida, UP-201307, India 
\par\end{center}
\begin{abstract}
It is shown that a realistic, controlled bidirectional remote state preparation
is possible using a large class of entangled quantum states having
a particular structure. Existing protocols of probabilistic, deterministic
and joint remote state preparation are generalized to obtain the corresponding
protocols of controlled bidirectional remote state preparation (CBRSP).
A general way of incorporating the effects of two well known noise processes, the 
amplitude-damping and phase-damping noise,
on the probabilistic CBRSP process is studied in detail by considering
that  noise only affects the travel qubits of the quantum channel
used for the probabilistic CBRSP process. Also indicated is how to account for
the effect of these noise channels on deterministic and joint remote state CBRSP protocols. 
\end{abstract}
Keywords: Remote state preparation, controlled bidirectional communication,
quantum communication, amplitude-damping noise, phase-damping noise.

\section{Introduction}

The concept of quantum teleportation was introduced by Bennett\emph{
}\textit{\emph{et al}}\emph{.} in 1993 \cite{Bennett}. In teleportation
an unknown quantum state is transmitted from a sender (Alice) to a receiver
(Bob) by using a shared entanglement and two bits of classical communication.
In this process, the unknown quantum state does not travel through
the quantum channel and thus this process does not have any classical
analogue. This extremely interesting nonclassical nature of the quantum
teleportation phenomenon drew considerable attention from the quantum
communication community and consequently a large number of modified
teleportation schemes have been proposed (for a review see Chapter
7 of Ref. \cite{my book}). For example, a large number of proposals
have been made for quantum information splitting (QIS) or quantum
secret sharing (QSS) \cite{Hillery}, controlled teleportation (CT)
\cite{Ct,A.Pathak}, hierarchical quantum information splitting (HQIS)
\cite{hierarchical,Shukla}, remote state preparation (RSP) \cite{Pati-original,LO-RSP,Bennet-RSP}.
All these schemes may be viewed as modified teleportation protocols.
Considerable attention has been devoted to bidirectional controlled
quantum state teleportation (BCST) \cite{Zha,Zha II,bi-directional-ourpaper,7qubit,sixqubit1,six-qubit-3,sixqubiit2,six-qubit-4,Li,5-qubit-c-qsdc},
controlled remote state preparation \cite{CRSP-hwang} and joint remote
state preparation (JRSP) \cite{JRSP-cluster-like,JRSP with paasive receiver,j-rsp-multi},
among others.

The original scheme of quantum teleportation was a one-way scheme
in which Alice teleports an unknown qubit to Bob by using two bits
of classical communication and an entangled state already shared by
them. Subsequently, it was modified by Huelga \textit{et al}. \cite{J. A. Vaccaro,J. A. Vaccaro-1}
and others to obtain schemes for bidirectional state teleportation
(BST), where both Alice and Bob can simultaneously transmit unknown
qubits to each other. It was also established that BST is useful for
the implementation of nonlocal quantum gates or quantum remote controls.
Recently, BST schemes have been generalized to obtain a set of schemes
for bidirectional controlled state teleportation (BCST) \cite{Zha,Zha II,bi-directional-ourpaper,7qubit,sixqubit1,six-qubit-3,sixqubiit2,six-qubit-4,Li,5-qubit-c-qsdc}.
These BCST schemes are three party schemes where BST is possible provided
the supervisor/controller (Charlie) discloses his information (measurement
outcome). To be more precise, in a usual BST scheme, Alice and Bob
can simultaneously transmit unknown quantum states to each other and
reconstruct the state received by them without the help of a third
party (Charlie), whereas in a BCST scheme also Alice and Bob can simultaneously
transmit unknown quantum states to each other as in BST, but could
not reconstruct the state received by them until the supervisor Charlie
(third party) allows them to do so.

In another line of research, conventional teleportation schemes were
modified to address a specific scenario, where the sender (Alice)
knows the state to be teleported, whereas the receiver (Bob) is completely
unaware of it. Such modified teleportation schemes are referred to
as schemes for the RSP. Thus, an RSP scheme may be viewed as a scheme
for teleportation of a known quantum state. RSP is possible with different
kinds of quantum channels and it's not limited to the remote preparation
of known single qubit states. In fact, several schemes for remote
preparation of known multipartite quantum states are also reported
in the recent past. Initially, the RSP scheme was designed for a single
qubit state \cite{Pati-original}. The scheme was probabilistic, but
it was shown that the task of remote preparation of a known single
qubit can be performed using a bit of classical communication and
a shared entanglement. Subsequently, this idea was generalized to
a deterministic RSP \cite{ba-an-deterministic}, wherein the remote
state could be prepared with unit success probability. However, the
reduction in classical communication with respect to a quantum teleportation
scheme that was achieved in the original scheme was lost here. The
idea of RSP has been generalized, recently, in many other ways. For
example, RSP schemes are proposed for remote preparation of a 4-qubit
$GHZ$ state \cite{RSP-1-GHZ}, multi-qubit $GHZ$ state \cite{RSP-multiparitite-ghz},
4-qubit cluster-type state \cite{RSP-4-qubit-cluster-type,RSP-cluster-type},
arbitrary two qubit state \cite{RSP-arbitrary-bipartite-usingGHZ-type}
and W state \cite{JRSP-W,JRSP-W-state}. Further, schemes for JRSP
have been constructed. A JRSP scheme involves at least three parties.
The simplest JRSP designed by An \cite{Ba An -RSP} can be described
as follows. Sender1, Sender2  and receiver 
share a $GHZ$ state. Sender1 and Sender2 want to jointly transmit
a qubit $|\psi\rangle=a|0\rangle+b\exp(i\phi)|1\rangle$ to the receiver,
but Sender1 knows $(a,\, b)$ and Sender2 knows $\phi$. In such a
case Sender1 measures his/her qubit using $\{|u_{0}\rangle=a|0\rangle+b|1\rangle$,
$|u_{1}\rangle=b|0\rangle-a|1\rangle\}$ basis set and announces the
result. Subsequently, Sender2 measures his/her qubit using $\{|v_{0}\rangle=\frac{|0\rangle+\exp(i\phi)|1\rangle}{\sqrt{2}},$
$|v_{1}\rangle=\frac{\exp(-i\phi)|0\rangle-|1\rangle}{\sqrt{2}}\}$
basis set (depending upon Sender1's measurement outcome Sender2 may
have to apply a unitary operation before measuring his/her qubit in $\{|v_{0}\rangle,  |v_{1}\rangle\}$  basis) and announces
the result. Using measurement outcomes of the senders, the receiver
can apply appropriate unitary operations and reconstruct the unknown
quantum state, which is jointly transmitted by the senders. Extending
this idea JRSP schemes were proposed for, among others, W and W-type
states \cite{JRSP-W,JRSP-W-state}, arbitrary two qubit state \cite{RSP-arbitrary-bipartite-usingGHZ-type}
and 4-qubit cluster like state \cite{JRSP-cluster-like}. Further
generalizing these ideas, schemes were proposed for controlled RSP
\cite{crsp1}, controlled JRSP \cite{cjrsp-1}, multi-party controlled
JRSP \cite{j-rsp-multi}, deterministic JRSP \cite{JRSP-BA-An} and
JRSP with a passive receiver \cite{JRSP with paasive receiver}. Thus,
a large number of variants of the RSP scheme have been investigated
in a one-directional scenario. However, except a recent work by Cao
and An \cite{ba-An-remote-state}, no serious effort has yet been
made to realize these features of RSP in bidirectional scenarios.
Such a scenario is practically relevant and can be visualized easily
in analogy with BCST as follows. Consider that both Alice and Bob
wish to remotely prepare quantum states that are known to the senders,
but unknown to the receivers and receivers can reconstruct the state
only when the supervisor Charlie allows them to do so. A scheme for
realizing this task would be referred to as the controlled bidirectional
remote state preparation (CBRSP). Such a scheme of CBRSP has been
proposed in \cite{ba-An-remote-state} by using a 5-qubit state \begin{equation}
|Q\rangle_{_{A_{1}A_{2}B_{1}B_{2}C_{1}}}=\frac{1}{2}\left(|00000\rangle+|01011\rangle+|10101\rangle+|11110\rangle\right){}_{A_{1}A_{2}B_{1}B_{2}C_{1}},\label{eq:Ba-An1}\end{equation}
 where the subscripts $A$, $B$ and $C$ indicate the qubits of Alice
(Sender), Bob (Receiver) and Charlie (Controller), respectively. For
our convenience, we can replace $A$, $B$ and $C$ by $S,$ $R,$
and $C,$ respectively, and after particle swapping rewrite the above
state as \begin{equation}
|Q\rangle_{_{S_{1}R_{1}S_{2}R_{2}C_{1}}}=\frac{1}{\sqrt{2}}\left(|\psi^{+}\rangle_{S_{1}R_{1}}|\psi^{+}\rangle_{S_{2}R_{2}}|+\rangle_{C_{1}}+|\psi^{-}\rangle_{S_{1}R_{1}}|\psi^{-}\rangle_{S_{2}R_{2}}|-\rangle_{C_{1}}\right),\label{eq:Ba-An2}\end{equation}
 where $|\psi^{\pm}\rangle=\frac{|00\rangle\pm|11\rangle}{\sqrt{2}}.$
Clearly, this is one of the states shown to be useful for BCST in
\cite{bi-directional-ourpaper}. To be precise, in Ref. \cite{bi-directional-ourpaper},
it was argued that the general form of the states that are useful
for BCST may be described as \begin{equation}
|\psi\rangle_{S_{1}R_{1}S_{2}R_{2}C_{1}}=\frac{1}{\sqrt{2}}\left(|\psi_{1}\rangle_{S_{1}R_{1}}|\psi_{2}\rangle_{S_{2}R_{2}}|a\rangle_{C_{1}}\pm|\psi_{3}\rangle_{S_{1}R_{1}}|\psi_{4}\rangle_{S_{2}R_{2}}|b\rangle_{C_{1}}\right),\label{eq:the 5-qubit state}\end{equation}
 where quantum states $|a\rangle$ and $|b\rangle$ satisfy $\langle a|b\rangle=\delta_{a,b}$,
$|\psi_{i}\rangle\in\left\{ |\psi^{+}\rangle,|\psi^{-}\rangle,|\phi^{+}\rangle,|\phi^{-}\rangle:|\psi_{1}\rangle\neq|\psi_{3}\rangle,|\psi_{2}\rangle\neq|\psi_{4}\rangle\right\} $,
$|\psi^{\pm}\rangle=\frac{|00\rangle\pm|11\rangle}{\sqrt{2}},$ $|\phi^{\pm}\rangle=\frac{|01\rangle\pm|10\rangle}{\sqrt{2}}$
and as before $S,$ $R,$ and $C$ represent sender, receiver and
controller%
\footnote{It is sufficient to consider $|a\rangle$ and $|b\rangle$ as single
qubit states, but it is not essential. One may consider them as 2-qubit
in particular or n-qubit (n>1) states in general, but that would only
introduce additional complexity.%
}. Here $|\psi_{i}\rangle$ is a Bell state and the condition \begin{equation}
|\psi_{1}\rangle\neq|\psi_{3}\rangle,|\psi_{2}\rangle\neq|\psi_{4}\rangle\label{eq:condition}\end{equation}
 ensures that Charlie's qubit is appropriately entangled with the
remaining 4 qubits \cite{bi-directional-ourpaper}. To be precise,
the receiver and the sender are unaware of the entangled (Bell) states
they share unless the controller measures his qubit in $\{|a\rangle,|b\rangle\}$
basis and discloses the result. We know that a shared Bell state and
two bits of classical communication is sufficient for teleportation,
but without the disclosure of the controller, the receiver will not
be able to choose appropriate unitary operations that will be required
to reconstruct the state. In what follows, we will extend our earlier
work on BCST \cite{bi-directional-ourpaper} and Cao and An's recent
work on CBRSP \cite{ba-An-remote-state} to show that if the state
(\ref{eq:the 5-qubit state}) satisfies the condition (\ref{eq:condition})
then on the disclosure of the outcome of Charlie's measurement in
$\{|a\rangle,|b\rangle\}$ basis, Alice and Bob will know with certainty
which two Bell states they share and consequently they will be able
to implement a scheme for probabilistic RSP or deterministic RSP in a controlled bidirectional manner.

In the discussion so far we have seen that probabilistic and deterministic
RSP, JRSP, controlled RSP and controlled JRSP are studied in detail
for one directional cases. We have also noted that RSP and teleportation
are closely linked phenomena as RSP can be viewed as teleportation
of a known qubit. Interestingly, a set of schemes for BCST have been
proposed, but only a single deterministic scheme for CBRSP using 5-qubit cluster
states is proposed till date. Neither any scheme for probabilistic
CBRSP, nor a scheme for controlled joint BRSP (CJBRSP) is proposed
until now. Keeping, these facts in mind, the present paper aims to
provide schemes for (i) probabilistic CBRSP, (ii) deterministic CBRSP
and (iii) deterministic CJBRSP using quantum states of the generic
form (\ref{eq:the 5-qubit state}). Further, we aim to show that the
schemes for probabilistic CBRSP, deterministic CBRSP and deterministic
CJBRSP can be realized using infinitely many different quantum channels
as the state described by (\ref{eq:the 5-qubit state}) can be constructed
in 144 different ways for each choice of $\{|a\rangle,|b\rangle\}$
basis and $\{|a\rangle,|b\rangle\}$ can be chosen in infinitely many
ways \cite{bi-directional-ourpaper}. Thus, the recently proposed
Cao and An scheme of deterministic CBRSP is just a special case of
infinitely many possibilities and their proposal, which is limited
to deterministic CBRSP can be extended to probabilistic CBRSP and
deterministic CJBRSP, too.

A practical implementation of any RSP protocol would imply taking into account the effect of the
ambient environment on the basic processes that constitute the protocol. For example, all control
protocols, as discussed here, depend upon the information sent to the sender, receiver by the controller.
This information is encoded in the travel qubits, i.e., the qubits prepared by
the controller and sent to different receivers and senders. These travel qubits traverse through physical space
and would be susceptible to the influence of the ambient environment, resulting in noise. Here we model
such a situation by studying the effect of two well known noise models, viz., the amplitude-damping and phase-damping
channels \cite{turchette, srikgp, ghoshqnd} on a probabilistic CBRSP protocol, and indicate how these effects can be
taken into account for a deterministic CBRSP as well as for deterministic CJBRSP. 

The paper is organized as follows. In Section \ref{sec:Probabilistic-controlled-bidirectional},
we propose a protocol of probabilistic CBRSP using quantum states
of the form (\ref{eq:the 5-qubit state}). In Section \ref{sec:Deterministic-controlled-bidirectional},
we show that the same states can be used to realize deterministic
CBRSP. We propose a protocol of deterministic CJBRSP, in Section \ref{sec:Deterministic-controlled-joint}.
In Section \ref{sec:Effect-of-damping}, we discuss the effect of the
amplitude-damping and the phase-damping noise channels on the probabilistic CBRSP scheme
described in this paper. Finally, we conclude in Section \ref{sec:Conclusion}.

\section{Probabilistic controlled bidirectional remote state preparation\label{sec:Probabilistic-controlled-bidirectional}}

Let us assume that the sender (S) and receiver (R) share an entangled
state $|\phi^{-}\rangle_{SR}=\left(\frac{|01\rangle-|10\rangle}{\sqrt{2}}\right)_{SR}$.
Now the sender wishes to transmit a known qubit $|\psi\rangle=a|0\rangle+b\exp(i\phi)|1\rangle$
to the receiver. Here, the sender knows the values of $a,\, b$ and
$\phi$. However, the receiver is completely unaware of these values.

Now we introduce a new basis set $\{|q_{1}\rangle,|q_{2}\rangle\}$,
where \begin{equation}
\begin{array}{lcl}
|q_{1}\rangle & = & a|0\rangle+b\exp(i\phi)|1\rangle=|\psi\rangle,\\
|q_{2}\rangle & = & b\exp(-i\phi)|0\rangle-a|1\rangle=U_{RSP}|\psi\rangle.\end{array}\label{eq:rsp1}\end{equation}
 Using $\{|q_{1}\rangle,|q_{2}\rangle\}$ basis set we can write \begin{equation}
\begin{array}{lcl}
|0\rangle & = & a|q_{1}\rangle+b\exp(i\phi)|q_{2}\rangle,\\
|1\rangle & = & b\exp(-i\phi)|q_{1}\rangle-a|q_{2}\rangle,\end{array}\label{eq:rsp2}\end{equation}
and using (\ref{eq:rsp2}) we can express the entangled state $|\phi^{-}\rangle_{SR}=\left(\frac{|01\rangle-|10\rangle}{\sqrt{2}}\right)_{SR}$
shared by the receiver and sender as follows: \begin{equation}
\begin{array}{lcl}
|\phi^{-}\rangle_{SR} & = & \frac{1}{\sqrt{2}}\left(|0\rangle_{S}|1\rangle_{R}-|1\rangle_{S}|0\rangle_{R}\right)\\
 & = & \frac{1}{\sqrt{2}}\left(\left(a|q_{1}\rangle+b\exp(i\phi)|q_{2}\rangle\right)_{S}|1\rangle_{R}-\left(b\exp(-i\phi)|q_{1}\rangle-a|q_{2}\rangle\right)_{S}\left.|0\rangle_{R}\right)\right)\\
 & = & \frac{1}{\sqrt{2}}|q_{1}\rangle_{S}\left(a|1\rangle-b\exp(-i\phi)|0\rangle\right)_{R}+\frac{1}{\sqrt{2}}|q_{2}\rangle_{S}\left(a|0\rangle+b\exp(i\phi)|1\rangle\right)_{R}\\
 & = & \frac{1}{\sqrt{2}}\left(|q_{2}q_{1}\rangle_{SR}-|q_{1}q_{2}\rangle_{SR}\right).\end{array}\label{eq:rsp3}\end{equation}
Now the sender measures his/her qubit in $\{|q_{1}\rangle,|q_{2}\rangle\}$
basis set and communicates the result to the receiver. From (\ref{eq:rsp3})
it is clear that if the sender's measurement outcome is $|q_{2}\rangle$
then the receiver does not need to do anything to reconstruct the
qubit unknown to him, but if the outcome of the sender's measurement
is $|q_{1}\rangle$ then the protocol fails as without the knowledge
of $\phi,$ the receiver will not be able to transform $|q_{2}\rangle$
into $|q_{1}\rangle=|\psi\rangle$. Thus, if the sender and receiver
share a prior entanglement, then the teleportation of a known qubit
requires only one projective measurement and one bit of classical
communication. Therefore, probabilistic remote state preparation requires
fewer resources than conventional teleportation. Now we may note that
RSP is also possible when other Bell states are used as initially
shared entanglement. To be precise, in $\{|q_{1}\rangle,|q_{2}\rangle\}$
basis, we can write \begin{equation}
\begin{array}{lcl}
|\phi^{+}\rangle_{SR} & = & \frac{1}{\sqrt{2}}\left(|0\rangle_{S}|1\rangle_{R}+|1\rangle_{S}|0\rangle_{R}\right)=\frac{1}{\sqrt{2}}Z_{R}\left(-|q_{2}q_{1}\rangle_{SR}+|q_{1}q_{2}\rangle_{SR}\right),\\
|\psi^{+}\rangle_{SR} & = & \frac{1}{\sqrt{2}}\left(|0\rangle_{S}|0\rangle_{R}+|1\rangle_{S}|1\rangle_{R}\right)=\frac{1}{\sqrt{2}}iY_{R}\left(-|q_{2}q_{1}\rangle_{SR}+|q_{1}q_{2}\rangle_{SR}\right),\\
|\psi^{-}\rangle_{SR} & = & \frac{1}{\sqrt{2}}\left(|0\rangle_{S}|0\rangle_{R}-|1\rangle_{S}|1\rangle_{R}\right)=\frac{1}{\sqrt{2}}X_{R}\left(|q_{2}q_{1}\rangle_{SR}-|q_{1}q_{2}\rangle_{SR}\right).\end{array}\label{eq:RSP4}\end{equation}
Thus, if we consider that the sender and receiver share a Bell state;
to prepare a remote state at the receiver's end, sender measures his/her
qubit in $\left\{ |q_{1}\rangle,|q_{2}\rangle\right\} $ basis and
announces the outcome, then the receiver can reconstruct the state
(in those cases where the output of the sender's measurement is $|q_{2}\rangle$)
by applying a unitary operator on his/her (receiver's) qubit, provided
the receiver knows which Bell state he/she shared with the sender.
The specific relation between the initially shared entangled state and
receiver's operation can be obtained from Eq. (\ref{eq:rsp3})-(\ref{eq:RSP4})
and the same is summarized in Table \ref{tab:probabilistic RSP}.
Now, if we consider that Charlie prepares the $5$-qubit state (\ref{eq:the 5-qubit state})
and sends qubits $S_{1},R_{2}$ to Alice and $R_{1},S_{2}$ to Bob.
In this situation Alice and Bob share two Bell states that can be
used for bidirectional remote state preparation (say the first Bell
state is shared for Alice to Bob transmission and the second one for
Bob to Alice transmission). Both of the senders can make the necessary
measurements in the rotated bases. To be precise, if Alice wishes
to remotely prepare $|\psi\rangle=a|0\rangle+b\exp(i\phi)|1\rangle$
at Bob's end, then she should measure $S_{1}$ qubit in $\left\{ |q_{1}\rangle,|q_{2}\rangle\right\} $
basis as given in (\ref{eq:rsp1}). Similarly, if Bob wishes to remotely
prepare $|\psi\rangle=a^{\prime}|0\rangle+b^{\prime}\exp(i\phi)|1\rangle$
at Alice's end, he should measure his qubit $S_{2}$ in $\left\{ |q_{1}^{\prime}\rangle,|q_{2}^{\prime}\rangle\right\} $
basis where $|q_{1}^{\prime}\rangle=a^{\prime}|0\rangle+b^{\prime}\exp(i\phi^{\prime})|1\rangle$
and $|q_{2}^{\prime}\rangle=b^{\prime}\exp(-i\phi^{\prime})|0\rangle-a^{\prime}|1\rangle$.
However, the receivers will not be able to apply the required unitary
operator unless they know which Bell states were initially shared
by them. For this information, they have to wait for Charlie's announcement
of a measurement outcome which he obtains by measuring his qubit in
$\left\{ |a\rangle,|b\rangle\right\} $ basis. Specifically, from
(\ref{eq:the 5-qubit state}) we can see that if Charlie's measurement
outcome is $|a\rangle$ then Alice to Bob (Bob to Alice) RSP channel
is $|\psi_{1}\rangle$ ($|\psi_{2}\rangle)$. Similarly, if Charlie's
measurement outcome is $|b\rangle$ then Alice to Bob (Bob to Alice)
RSP channel is $|\psi_{3}\rangle$ ($|\psi_{4}\rangle)$. Thus, Charlie
can control the bidirectional remote state preparation protocol described
here. Further, since the protocol succeeds only when the sender's
measurement outcome is $|q_{2}\rangle$ ($|q_{2}^{\prime}\rangle$),
thus the protocol is of probabilistic nature. In brief, we have obtained
a generalized Pati-type scheme for probabilistic CBRSP.

\begin{table}
\begin{centering}
\begin{tabular}{|>{\centering}p{1.5in}|>{\centering}p{1in}|>{\centering}p{1in}|>{\centering}p{1in}|>{\centering}p{1in}|}
\hline 
 & \multicolumn{4}{c|}{Initial state shared by the sender and receiver }\tabularnewline
\hline 
 & $|\psi^{+}\rangle$  & $|\psi^{-}\rangle$  & $|\phi^{+}\rangle$  & $|\phi^{-}\rangle$\tabularnewline
\hline 
Sender's measurement outcome in $\{|q_{1}\rangle,|q_{2}\rangle\}$
basis  & Receiver's operation  & Receiver's operation  & Receiver's operation  & Receiver's operation \tabularnewline
\hline 
$|q_{2}\rangle$  & $iY$  & $X$  & $Z$  & $I$\tabularnewline
\hline 
$|q_{1}\rangle$  & \multicolumn{4}{c|}{Protocol fails}\tabularnewline
\hline
\end{tabular}
\par\end{centering}

\caption{\label{tab:probabilistic RSP} Relation between the measurement outcomes
of the sender and the unitary operations applied by the receiver to
implement probabilistic remote state preparation using different initial
states. }
\end{table}

\section{Deterministic controlled bidirectional remote state preparation \label{sec:Deterministic-controlled-bidirectional}}

The CBRSP scheme described in the previous section and its parent
scheme (one-directional Pati scheme) was probabilistic. In \cite{ba-an-deterministic}
the Pati scheme was generalized to obtain a one-directional deterministic
scheme. In their original scheme \cite{ba-an-deterministic}, the
sender (S) and receiver (R) start with a shared Bell state $|\psi^{+}\rangle=\frac{\left(|00\rangle+|11\rangle\right)_{SR}}{\sqrt{2}}$,
where the sender (S) has the first qubit and the receiver (R) has
the second qubit. The sender also prepares another ancillary qubit
in state $|0\rangle$ which is indexed with the subscript $S^{\prime}.$
Subsequently, the sender applies a CONT operation using the $S$ qubit
as the control qubit and the $S^{\prime}$ qubit as the target qubit
to obtain a combined state \begin{equation}
\begin{array}{c}
{\rm CNOT}\\
S\rightarrow S^{\prime}\end{array}\frac{\left(|00\rangle+|11\rangle\right)_{SR}}{\sqrt{2}}|0\rangle_{S^{\prime}}=\frac{\left(|000\rangle+|111\rangle\right)_{SS^{\prime}R}}{\sqrt{2}}={\rm GHZ^{0+}},\label{eq:GHZ-an}\end{equation}
which is nothing but a $GHZ$ state. For our convenience, we have
indexed the $GHZ$ states produced in this way with a superscript
$0+$, where $0$ is the decimal value of the first component of the
superposition that forms the $GHZ$ state (i.e., decimal value of
000) and the $+$ sign denotes the relative phase between the two
components of the superposition. Before, we describe the original
protocol of \cite{ba-an-deterministic} in detail, we would like to note a few things
for our convenience. Firstly, a Bell state can be expressed in general
as $|\psi\rangle_{{\rm Bell}}=\frac{|ij\rangle\pm|\bar{i}\bar{j}\rangle}{\sqrt{2}}$
with $i,j\in\left\{ 0,1\}\right\} $. Thus, \begin{equation}
\begin{array}{c}
{\rm CNOT}\\
S\rightarrow S^{\prime}\end{array}\frac{\left(|ij\rangle\pm|\bar{i}\bar{j}\rangle\right)_{{\rm SR}}}{\sqrt{2}}|0\rangle_{S^{\prime}}=\frac{\left(|iij\rangle\pm|\bar{i}\bar{i}\bar{j}\rangle\right)_{SS^{\prime}R}}{\sqrt{2}}={\rm GHZ^{x\pm}},\label{eq:vishal1}\end{equation}
where $x$ is the decimal value of binary number $iij$ and $\pm$
denotes the relative phase between the two components of the superposition.
This would lead to 4 $GHZ$ states depending upon the choice of initial
Bell state. Similarly, we can obtain 4 more $GHZ$ states if the sender
prepares the ancillary qubit in the state $|1\rangle$. Specifically,
the relation between the $GHZ$ states produced and the Bell state
used are as follows: \begin{equation}
\begin{array}{lclclcl}
\begin{array}{c}
{\rm CNOT}\\
S\rightarrow S^{\prime}\end{array}|\psi^{\pm}\rangle_{SR}|0\rangle_{S^{'}} & = & \begin{array}{c}
{\rm CNOT}\\
S\rightarrow S^{\prime}\end{array}\frac{\left(|00\rangle\pm|11\rangle\right)_{SR}}{\sqrt{2}}|0\rangle_{S^{'}} & = & {\rm GHZ^{0\pm}} & = & \frac{\left(|000\rangle\pm|111\rangle\right)_{SS^{\prime}R}}{\sqrt{2}},\\
\begin{array}{c}
{\rm CNOT}\\
S\rightarrow S^{\prime}\end{array}|\phi^{\pm}\rangle_{SR}|0\rangle_{S^{'}} & = & \begin{array}{c}
{\rm CNOT}\\
S\rightarrow S^{\prime}\end{array}\frac{\left(|01\rangle\pm|10\rangle\right)_{SR}}{\sqrt{2}}|0\rangle_{S^{'}} & = & {\rm GHZ^{1\pm}} & = & \frac{\left(|001\rangle\pm|110\rangle\right)_{SS^{\prime}R}}{\sqrt{2}},\\
\begin{array}{c}
{\rm CNOT}\\
S\rightarrow S^{\prime}\end{array}|\psi^{\pm}\rangle_{SR}|1\rangle_{S^{'}} & = & \begin{array}{c}
{\rm CNOT}\\
S\rightarrow S^{\prime}\end{array}\frac{\left(|00\rangle\pm|11\rangle\right)_{SR}}{\sqrt{2}}|1\rangle_{S^{'}} & = & {\rm GHZ^{2\pm}} & = & \frac{\left(|010\rangle\pm|101\rangle\right)_{SS^{\prime}R}}{\sqrt{2}},\\
\begin{array}{c}
{\rm CNOT}\\
S\rightarrow S^{\prime}\end{array}|\phi^{\pm}\rangle_{SR}|1\rangle_{S^{'}} & = & \begin{array}{c}
{\rm CNOT}\\
S\rightarrow S^{\prime}\end{array}\frac{\left(|01\rangle\pm|10\rangle\right)_{SR}}{\sqrt{2}}|1\rangle_{S^{'}} & = & {\rm GHZ^{3\pm}} & = & \frac{\left(|011\rangle\pm|100\rangle\right)_{SS^{\prime}R}}{\sqrt{2}}.\end{array}\label{eq:relation}\end{equation}
Now we continue with the original idea of \cite{ba-an-deterministic}.
After creating $GHZ$ state (\ref{eq:GHZ-an}), the sender (who wishes
to remotely prepare a quantum state $|\psi\rangle=a|0\rangle+b\exp(i\phi)|1\rangle$)
measures his/her first qubit in a new basis $\left\{ |u_{0}\rangle=a|0\rangle+b|1\rangle,|u_{1}\rangle=b|0\rangle-a|1\rangle\right\} $.
Using this, we can write $|0\rangle=a|u_{0}\rangle+b|u_{1}\rangle$
and $|1\rangle=b|u_{0}\rangle-a|u_{1}\rangle$ and consequently the
$GHZ$ state (\ref{eq:GHZ-an}) can be expressed as \begin{equation}
\begin{array}{lcl}
{\rm GHZ^{0+}} & = & \frac{\left(|000\rangle+|111\rangle\right)_{SS^{\prime}R}}{\sqrt{2}}\\
 & = & \frac{1}{\sqrt{2}}\left[|u_{0}\rangle\left(a|00\rangle+b|11\rangle\right)+|u_{1}\rangle\left(b|00\rangle-a|11\rangle\right)\right]_{SS^{\prime}R}\end{array}.\label{eq:psi1}\end{equation}
From Eq. (\ref{eq:psi1}) we can easily see that if sender's measurement
of the first qubit in $\left\{ |u_{0}\rangle,|u_{1}\rangle\right\} $
basis yields $|u_{0}\rangle$ then the state of the remaining qubits
collapses to $|\Psi_{0}\rangle_{S^{\prime}R}=\left(a|00\rangle+b|11\rangle\right)_{S^{\prime}R}$
and if the measurement yields $|u_{1}\rangle$ then the state of the
remaining qubits collapses to $|\Psi_{1}\rangle_{S^{\prime}R}=\left(b|00\rangle-a|11\rangle\right)_{S^{\prime}R}$.
In \cite{ba-an-deterministic}, the sender follows two different routes
depending on the outcome of the previous measurement. To be precise,
when he/she obtains $|u_{0}\rangle$ he/she applies a phase gate $\Pi=\left(\begin{array}{cc}
1 & 0\\
0 & \exp\left(2i\phi\right)\end{array}\right)$ on the qubit $S^{\prime}$ to transform $|\Psi_{0}\rangle_{S^{\prime}R}$
to \begin{equation}
|\Psi_{0}^{\prime}\rangle_{S^{\prime}R}=\Pi_{S^{\prime}}|\Psi_{0}\rangle_{S^{\prime}R}=\left(a|00\rangle+b\exp\left(2i\phi\right)|11\rangle\right)_{S^{\prime}R}.\label{eq:crsp1}\end{equation}
 However, if his/her measurement yields $|u_{1}\rangle$, then he/she
does nothing (i.e., keeps the state $|\Psi_{1}\rangle_{S^{\prime}R}=\left(b|00\rangle-a|11\rangle\right)_{S^{\prime}R}$
unchanged). Subsequently he/she measures the $S^{\prime}$ qubit in
$\left\{ |v_{0}\rangle=\frac{|0\rangle+\exp(i\phi)|1\rangle}{\sqrt{2}},|v_{1}\rangle=\frac{\exp(-i\phi)|0\rangle-|1\rangle}{\sqrt{2}}\right\} $
basis. By expressing $S^{\prime}$ qubit in $\left\{ |v_{0}\rangle,|v_{1}\rangle\right\} $
basis, the states $|\Psi_{0}^{\prime}\rangle_{S^{\prime}R}$ and $|\Psi_{1}\rangle_{S^{\prime}R}$
can be rewritten as follows \begin{equation}
|\Psi_{0}^{\prime}\rangle_{S^{\prime}R}=\frac{1}{\sqrt{2}}\left[|v_{0}\rangle_{S^{\prime}}\left(a|0\rangle+b\exp(i\phi)|1\rangle\right)_{R}+\exp(i\phi)|v_{1}\rangle_{S^{\prime}}\left(a|0\rangle-b\exp(i\phi)|1\rangle\right)_{R}\right],\label{eq:ccrsp2}\end{equation}
 and \begin{equation}
|\Psi_{1}\rangle_{S^{\prime}R}=\frac{1}{\sqrt{2}}\left[\exp(-i\phi)|v_{0}\rangle_{S^{\prime}}\left(b\exp(i\phi)|0\rangle-a|1\rangle\right)_{R}+|v_{1}\rangle_{S^{\prime}}\left(b\exp(i\phi)|0\rangle+a|1\rangle\right)_{R}\right].\label{eq:ccrsp3}\end{equation}
 From Eq. (\ref{eq:ccrsp2})-(\ref{eq:ccrsp3}) we can clearly conclude
that if the sender's measurements yield $|u_{0}\rangle|v_{0}\rangle,$
$|u_{0}\rangle|v_{1}\rangle$, $|u_{1}\rangle|v_{0}\rangle,$ and
$|u_{1}\rangle|v_{1}\rangle$ respectively, then the receiver can
reconstruct the transmitted state by applying $I,\, Z,\, iY,$ and
$X$ operators, respectively. Thus we have a protocol for deterministic
RSP using an initially shared Bell state which was prepared in $|\psi^{+}\rangle=\frac{\left(|00\rangle+|11\rangle\right)_{SR}}{\sqrt{2}}$.
It is easy to check that if the receiver and sender start from other
Bell states, then, also the above protocol succeeds in deterministic
RSP. However, the unitary operations to be performed by the receiver
are different for different initial shared states as shown in Table
\ref{tab:Deterministic-RSP}. Now, it is easy to observe that if Alice
and Bob start with a quantum state of the form (\ref{eq:the 5-qubit state})
and in a manner analogous to the probabilistic CBRSP described in
the previous section use one Bell state for Alice to Bob transmission
and the other one for Bob to Alice transmission, then both of them
will be able to remotely prepare their quantum states provided they
know which Bell states they share. Thus, only after Charlie measures
his qubit in $\left\{ |a\rangle,|b\rangle\right\} $ basis and announces
the result, Alice and Bob will know which operations from the Table
\ref{tab:Deterministic-RSP} are to be used. However, once Alice and
Bob know Charlie's measurement outcome, Alice (Bob) can deterministically
prepare her (his) quantum state at Bob's (Alice's) side. Thus all
states of the form (\ref{eq:the 5-qubit state}) would lead to deterministic
CBRSP. Now it appears obvious why a special case of (\ref{eq:the 5-qubit state})
used by Cao and An in Ref. \cite{ba-An-remote-state} lead to the
first ever protocol of CBRSP. Here we observe that there exists infinitely
many 5-qubit states that can be used for CBRSP and have provided a
general structure of those states as (\ref{eq:the 5-qubit state}).

\begin{table}
\begin{centering}
\begin{tabular}{|>{\centering}p{1in}|>{\centering}p{1in}|>{\centering}p{1in}|>{\centering}p{1in}|>{\centering}p{1in}|>{\centering}p{1in}|}
\hline 
\multicolumn{2}{|c|}{} & \multicolumn{4}{c|}{Initial state shared by the sender and receiver}\tabularnewline
\hline 
\multicolumn{2}{|c|}{} & $|\psi^{+}\rangle$  & $|\psi^{-}\rangle$  & $|\phi^{+}\rangle$  & $|\phi^{-}\rangle$\tabularnewline
\hline 
Sender's measurement outcome in $\{|u_{0}\rangle,|u_{1}\rangle\}$
basis  & Sender's measurement outcome in $\{|v_{0}\rangle,|v_{1}\rangle\}$
basis  & Receiver's operation  & Receiver's operation  & Receiver's operation  & Receiver's operation \tabularnewline
\hline 
$|u_{0}\rangle$  & $|v_{0}\rangle$  & $I$  & $Z$  & $X$  & $iY$\tabularnewline
\hline 
$|u_{0}\rangle$  & $|v_{1}\rangle$  & $Z$  & $I$  & $iY$  & $X$\tabularnewline
\hline 
$|u_{1}\rangle$  & $|v_{0}\rangle$  & $iY$  & $X$  & $Z$  & $I$\tabularnewline
\hline 
$|u_{1}\rangle$  & $|v_{1}\rangle$  & $X$  & $iY$  & $I$  & $Z$\tabularnewline
\hline
\end{tabular}
\par\end{centering}

\caption{\label{tab:Deterministic-RSP}Relation between the measurement outcomes
of the sender and the unitary operations applied by the receiver to
implement deterministic remote state preparation using different initial
states. }
\end{table}

\section{Deterministic controlled joint bidirectional remote state preparation
\label{sec:Deterministic-controlled-joint}}

In an usual JRSP scheme a quantum state $a|0\rangle+b\exp(i\phi)|1\rangle$
is jointly prepared at the receiver's end by two senders. One of the
senders knows the value of $a,\, b$ and the other one knows the value
of $\phi.$ Taking a careful look into the scheme presented in \cite{ba-an-deterministic}
of deterministic RSP, described above, one can quickly recognize that
the projective measurement performed on qubit $S$ (in $\left\{ |u_{0}\rangle,|u_{1}\rangle\right\} $
basis) only requires the knowledge of $a,\, b$, while application
of unitary operator $\Pi$ and measurement in $\left\{ |v_{0}\rangle,|v_{1}\rangle\right\} $
basis performed on the qubit $S^{\prime}$ only requires the knowledge
of $\phi$. Thus, two different parties can perform this operation.
Specifically, we may provide access of qubit $S$ to a person having
knowledge of $a,b$ and access of qubit $S^{\prime}$ to another person
having knowledge of $\phi$. To see that this is sufficient for JRSP,
consider that Sender1 prepares a $GHZ^{^{0+}}$ state in such a way
that he/she keeps the first qubit ($S)$, and sends the second ($S^{\prime}$)
and third ($R$ ) qubits to Sender2 and the receiver, respectively.
Subsequently, Sender1 measures his/her qubit in $\left\{ |u_{0}\rangle,|u_{1}\rangle\right\} $
basis and communicates the result to the Sender2 and the receiver.
Sender2 applies $\Pi$ gate on his/her qubit if required (i.e., iff
the outcome of Sender1's measurement is $|u_{0}\rangle)$ and measures
it in $\left\{ |v_{0}\rangle,|v_{1}\rangle\right\} $ basis and communicates
the result to the receiver. The receiver can use Table \ref{tab:Deterministic-RSP}
or equivalently Table \ref{tab:3} to apply an appropriate operation
to reconstruct $a|0\rangle+b\exp(i\phi)|1\rangle$ at his/her end.
Now it is not difficult to observe that by combining Eqs. (\ref{eq:the 5-qubit state})
and (\ref{eq:relation}) we can obtain \begin{equation}
\begin{array}{lcl}
 &  & \begin{array}{c}
{\rm CNOT}\\
S_{1}\rightarrow S_{1}^{\prime}\end{array}\begin{array}{c}
{\rm CNOT}\\
S_{2}\rightarrow S_{2}^{\prime}\end{array}|\psi\rangle_{S_{1}R_{1}S_{2}R_{2}C_{1}}|00\rangle_{S_{1}^{\prime}S_{2}^{\prime}}\\
 & = & \frac{1}{\sqrt{2}}\begin{array}{c}
{\rm CNOT}\\
S_{1}\rightarrow S_{1}^{\prime}\end{array}\begin{array}{c}
{\rm CNOT}\\
S_{2}\rightarrow S_{2}^{\prime}\end{array}\left(|\psi_{1}\rangle_{S_{1}R_{1}}|\psi_{2}\rangle_{S_{2}R_{2}}|a\rangle_{C_{1}}\pm|\psi_{3}\rangle_{S_{1}R_{1}}|\psi_{4}\rangle_{S_{2}R_{2}}|b\rangle_{C_{1}}\right)|00\rangle_{S_{1}^{\prime}S_{2}^{\prime}},\\
 & = & \frac{1}{\sqrt{2}}\left(|GHZ_{1}\rangle_{S_{1}S_{1}^{\prime}R_{1}}|GHZ_{2}\rangle_{S_{2}S_{2}^{\prime}R_{2}}|a\rangle_{C_{1}}\pm|GHZ_{3}\rangle_{S_{1}S_{1}^{\prime}R_{1}}|GHZ_{4}\rangle_{S_{2}S_{2}^{\prime}R_{2}}|b\rangle_{C_{1}}\right),\end{array}\label{eq:7-qubit-1}\end{equation}
with $GHZ_{1}\neq GHZ_{3}$ and $GHZ_{2}\neq GHZ_{4}$ and $GHZ_{i}\in\left\{ GHZ^{0\pm},GHZ^{1\pm}\right\} $
with $i\in\left\{ 1,2,3,4\right\} $. The 7-qubit state (\ref{eq:7-qubit-1})
that originated from our 5-qubit channel (\ref{eq:the 5-qubit state})
is clearly sufficient for the deterministic CJBRSP. As from Table \ref{tab:3} we can
see that without the knowledge of the $GHZ_{i}$ shared by Sender1,
Sender2 and the receiver, it will be impossible for the receiver to
reconstruct the quantum state transmitted. On disclosure of controller's
measurement, shared states reduce to the product of two $GHZ$ states.
One of them can be used for JRSP in one direction and the other for
the JRSP in the other direction. Following the same logic as followed
in \cite{bi-directional-ourpaper}, we can show that for each choice
of basis set $\left\{ |a\rangle,|b\rangle\right\} $ there are 144
alternative ways to satisfy the condition $GHZ_{1}\neq GHZ_{3}$ and
$GHZ_{2}\neq GHZ_{4}$ and $GHZ_{i}\in\left\{ GHZ^{0\pm},GHZ^{1\pm}\right\} $
with $i\in\left\{ 1,2,3,4\right\} $ (without $\pm$ sign) and thus
to construct alternative quantum channels of the form (\ref{eq:7-qubit-1})
(cf. \cite{bi-directional-ourpaper}). Interestingly, (\ref{eq:7-qubit-1})
does not exhaust all the possibilities. For example, we can think
of quantum states of the form \begin{equation}
\begin{array}{c}
{\rm CNOT}\\
S_{1}\rightarrow S_{1}^{\prime}\end{array}\begin{array}{c}
{\rm CNOT}\\
S_{2}\rightarrow S_{2}^{\prime}\end{array}|\psi\rangle_{S_{1}R_{1}S_{2}R_{2}C_{1}}|11\rangle_{S_{1}^{\prime}S_{2}^{\prime}}=\frac{1}{\sqrt{2}}\left(|GHZ_{1}\rangle_{S_{1}S_{1}^{\prime}R_{1}}|GHZ_{2}\rangle_{S_{2}S_{2}^{\prime}R_{2}}|a\rangle_{C_{1}}\pm|GHZ_{3}\rangle_{S_{1}S_{1}^{\prime}R_{1}}|GHZ_{4}\rangle_{S_{2}S_{2}^{\prime}R_{2}}|b\rangle_{C_{1}}\right)\label{eq:7-qubit-2}\end{equation}
with $GHZ_{1}\neq GHZ_{3}$ and $GHZ_{2}\neq GHZ_{4}$ and $GHZ_{i}\in\left\{ GHZ^{2\pm},GHZ^{3\pm}\right\} $
with $i\in\left\{ 1,2,3,4\right\} $. Once again we will obtain 144
alternative quantum channels for each choice of basis set $\left\{ |a\rangle,|b\rangle\right\} $.
All these states are also useful for CJBRSP. However, in this case
we have to slightly modify the intrinsic protocol of JRSP. To be
precise, if we assume that Sender1, Sender2 and receiver share a state
$|\psi\rangle_{SS^{\prime}R}=GHZ\in\left\{ GHZ^{2\pm},GHZ^{3\pm}\right\} $,
and measurement of Sender1 on $S$ using \{$|u_{0}\rangle,|u_{1}\rangle\}$
basis yields $|u_{0}\rangle$ ($|u_{1}\rangle$) then Sender2 does
nothing (applies unitary operator $\Pi$) on the qubit $S^{\prime}$
before measuring the qubit using \{$|v_{0}\rangle,|v_{1}\rangle\}$
basis. Subsequently, using these measurement outcomes, the receiver
will be able to reconstruct the unknown state by applying appropriate
unitary operators described in Table \ref{tab:3}. Combining the above
two quantum channels (i.e., combining Eqns. (\ref{eq:7-qubit-1}) and (\ref{eq:7-qubit-2})) now we can also think of a very general quantum
channel for CJBRSP of the form \begin{equation}
|\psi\rangle=\frac{1}{\sqrt{2}}\left(|GHZ_{1}\rangle_{S_{1}S_{1}^{\prime}R_{1}}|GHZ_{2}\rangle_{S_{2}S_{2}^{\prime}R_{2}}|a\rangle_{C_{1}}\pm|GHZ_{3}\rangle_{S_{1}S_{1}^{\prime}R_{1}}|GHZ_{4}\rangle_{S_{2}S_{2}^{\prime}R_{2}}|b\rangle_{C_{1}}\right)\label{eq:7-qubit-3}\end{equation}
 with $GHZ_{1}\neq GHZ_{3}$ and $GHZ_{2}\neq GHZ_{4}$ and $GHZ_{i}\in\left\{ GHZ^{0\pm},GHZ^{1\pm},GHZ^{2\pm},GHZ^{3\pm}\right\} $
with $i\in\left\{ 1,2,3,4\right\} $. Such a state will obviously
work as a quantum channel for CJBRSP. To be precise, in a particular
implementation of the scheme all parties know $GHZ_{1}, GHZ_{2}, GHZ_{3}, GHZ_{4},$
and Sender2 can always apply his/her operation; however in the most
general case (for example, consider that $GHZ_{1}=GHZ^{1+}$ and $GHZ_{3}=GHZ^{3+}$)
Sender2 may have to wait till the disclosure of the controller to
decide in which case he/she will apply the $\Pi$ operation.

\begin{table}
\begin{tabular}{|>{\centering}p{1in}|>{\centering}p{1in}|>{\centering}p{1in}|>{\centering}p{1in}|>{\centering}p{1in}|>{\centering}p{1in}|}
\hline 
\multicolumn{2}{|c|}{} & \multicolumn{4}{c|}{Initial state shared by Sender1, Sender2 and receiver}\tabularnewline
\hline 
\multicolumn{2}{|c|}{} & $GHZ^{0+}$ or $GHZ^{2+}$  & $GHZ^{0-}$ or $GHZ^{2-}$  & $GHZ^{1+}$ or $GHZ^{3+}$  & $GHZ^{1-}$ or $GHZ^{3-}$\tabularnewline
\hline 
Sender1's measurement outcome in $\{|u_{0}\rangle,|u_{1}\rangle\}$
basis  & Sender2's measurement outcome in $\{|v_{0}\rangle,|v_{1}\rangle\}$
basis  & Receiver's operation  & Receiver's operation  & Receiver's operation  & Receiver's operation \tabularnewline
\hline 
$|u_{0}\rangle$  & $|v_{0}\rangle$  & $I$  & $Z$  & $X$  & $iY$\tabularnewline
\hline 
$|u_{0}\rangle$  & $|v_{1}\rangle$  & $Z$  & $I$  & $iY$  & $X$\tabularnewline
\hline 
$|u_{1}\rangle$  & $|v_{0}\rangle$  & $iY$  & $X$  & $Z$  & $I$\tabularnewline
\hline 
$|u_{1}\rangle$  & $|v_{1}\rangle$  & $X$  & $iY$  & $I$  & $Z$\tabularnewline
\hline
\end{tabular}\caption{\label{tab:3}Table for reconstruction of the quantum state for JRSP.
Protocol to be followed for shared states of the form $GHZ_{i}\in\left\{ GHZ^{0\pm},GHZ^{1\pm}\right\} $
is slightly different from that for the shared state of the form $GHZ_{i}\in\left\{ GHZ^{2\pm},GHZ^{3\pm}\right\} $. }
\end{table}

\section{Effect of the amplitude-damping noise and the phase-damping noise
on the CBRSP process \label{sec:Effect-of-damping}}

In this section, we consider the effect of noise on remotely prepared
quantum states when the travel qubits  pass through either
the amplitude-damping noisy environment or the phase-damping noisy
environment. The amplitude-damping noise model is characterized
by the following Kraus operators \cite{RSP-with-noise}  \begin{equation}
E_{0}^{A}=\left[\begin{array}{cc}
1 & 0\\
0 & \sqrt{1-\eta_{A}}\end{array}\right],\,\,\,\,\,\,\,\,\,\,\,\,\,\,\, E_{1}^{A}=\left[\begin{array}{cc}
0 & \sqrt{\eta_{A}}\\
0 & 0\end{array}\right],\label{eq:Krauss-amp-damping}\end{equation}
where $\eta_{A}$ ($0\leq\eta_{A}\leq1$)
describes the probability of error due to amplitude-damping noisy
environment when a travel qubit pass through it. $\eta_{A}$ is also
referred to as decoherence rate. Similarly, phase-damping noise model
is characterized by the following Kraus operators \begin{equation}
E_{0}^{P}=\sqrt{1-\eta_{P}}\left[\begin{array}{cc}
1 & 0\\
0 & 1\end{array}\right],\,\,\,\,\,\,\,\,\,\,\,\,\,\,\, E_{1}^{P}=\sqrt{\eta_{P}}\left[\begin{array}{cc}
1 & 0\\
0 & 0\end{array}\right],\,\,\,\,\,\,\,\,\,\,\,\,\,\,\, E_{2}^{P}=\sqrt{\eta_{P}}\left[\begin{array}{cc}
0 & 0\\
0 & 1\end{array}\right],\label{eq:Krauss-phase-damping}\end{equation}
where $\eta_{P}$ ($0\leq\eta_{P}\leq1$)
is the decoherence rate for the phase-damping noise. 

In Section \ref{sec:Probabilistic-controlled-bidirectional}, we have
proposed a scheme for probabilistic CBRSP using a 5-qubit quantum
channel $|\psi\rangle_{S_{1}R_{1}S_{2}R_{2}C_{1}}$ having a general
form described by (\ref{eq:the 5-qubit state}). As the state is
pure it is straightforward to obtain the corresponding density matrix
\[
\rho=|\psi\rangle_{S_{1}R_{1}S_{2}R_{2}C_{1}}{}_{S_{1}R_{1}S_{2}R_{2}C_{1}}\langle\psi|.\]
Now the effect of the noisy environment described by (\ref{eq:Krauss-amp-damping})
or (\ref{eq:Krauss-phase-damping}) on the density operator $\rho$
is \begin{equation}
\rho_{k}=\sum_{i,j}E_{i,S_{1}}^{k}\otimes E_{j,R_{1}}^{k}\otimes E_{j,S_{2}}^{k}\otimes E_{i,R_{2}}^{k}\otimes I_{2,C_{1}}\rho\left(E_{i,S_{1}}^{k}\otimes E_{j,R_{1}}^{k}\otimes E_{j,S_{2}}^{k}\otimes E_{i,R_{2}}^{k}\otimes I_{2,C_{1}}\right)^{\dagger},\label{eq:noise-effected-density-matrix}\end{equation}
where $I_{2}$ is a $2\times2$ identity matrix, $k\in\{A,P\}$. For
for $k=A,$ i.e., for amplitude-damping noise $i,j\in\{1,2\}$, while for
$k=P$, i.e., for phase-damping noise $i,j\in\{1,2,3\}$, and the second
subscripts on the Kraus operators are included to specify the specific
qubit on which it is to be operated. In the construction of (\ref{eq:noise-effected-density-matrix})
we have considered that the qubit of the controller ($C_{1}$) is
not affected by the noise as it is not transmitted through the noisy environment.
Further, it is assumed that both the qubits sent to Alice (i.e., $S_{1}$
and $R_{2}$ qubits) are affected by the same Kraus operator and similarly,
the qubits $R_{1}$ and $S_{2}$ sent to Bob are also affected by
the same Kraus operator. As a consequence of the noisy environment,
the initial quantum channel which was pure gets transformed into a
mixed state. Senders and receivers faithfully apply the probabilistic
CBRSP scheme on $\rho_{k}$. To be precise, we assume that Alice (Sender1)
and Bob (Sender2) wish to remotely prepare qubits $a_{1}|0\rangle+b_{1}\exp(i\phi_{1})|1\rangle$
and $a_{2}|0\rangle+b_{2}\exp(i\phi_{2})|1\rangle$ at the side of
Bob (Receiver1) and Alice (Receiver2), respectively. To do so in accordance
with the probabilistic CBRSP scheme described in the present work, $S_{1}$ qubit
is measured by Alice using $\left\{ |q_{1}\rangle_{S_{1}}=a_{1}|0\rangle+b_{1}\exp(i\phi_{1})|1\rangle,\,|q_{2}\rangle_{S_{1}}=b_{1}\exp(-i\phi_{1})|0\rangle-a_{1}|1\rangle\right\} $
basis, $S_{2}$ qubit is measured by Bob using $\left\{ |q_{1}\rangle_{S_{2}}=a_{2}|0\rangle+b_{2}\exp(i\phi_{2})|1\rangle,\,|q_{2}\rangle_{S_{2}}=b_{2}\exp(-i\phi_{2})|0\rangle-a_{2}|1\rangle\right\} $,
and $C_{1}$ qubit is measured by Charlie\textbackslash{}controller
using $\left\{ |a\rangle,|b\rangle\right\} $ basis. As in a probabilistic
CBRSP scheme, the RSP in a specific direction succeeds only when corresponding
senders measurement yields $|q_{2}\rangle$. For a successful probabilistic
CBRSP, measurements on $S_{1}$ and $S_{2}$ should yield $|q_{2}\rangle_{S_{1}}$
and $|q_{2}\rangle_{S_{2}}$, respectively. For our convenience, we
assume that the measurement of controller yields $|b\rangle.$ Thus,
to selectively choose these outcomes we have to apply the operator \[
U=\left(|q_{2}\rangle_{S_{1}S_{1}}\langle q_{2}|\right)\otimes I_{2,R_{1}}\left(|q_{2}\rangle_{S_{2}S_{2}}\langle q_{2}|\right)\otimes I_{2,R_{2}}\otimes|b\rangle_{C_{1}C_{1}}\langle b|\]
on $\rho_{k}$ yielding an unnormalized quantum state \[
\rho_{k_{1}}=U\rho_{k}U^{\dagger},\]
which can be normalized to yield a quantum state \[
\rho_{k_{2}}=\frac{\rho_{k_{1}}}{{\rm Tr}\left(\rho_{k_{1}}\right)}.\]
Now the combined states of the qubits $R_{1}$ and $R_{2}$ or $\rho_{k_{3}}$
can be obtained form $\rho_{k_{2}}$, by tracing out the qubits that
are already measured. Specifically, \[
\rho_{k_{3}}={\rm Tr}{}_{S_{1}S_{2}C_{1}}\left(\rho_{k_{2}}\right).\]
Depending upon the specific choice of the initial quantum channel
we may have to apply appropriate Pauli operators on $\rho_{k_{3}}$
to obtain the final quantum state $\rho_{k,{\rm out}}$ which is the
product of the quantum states produced on the side of the Receivers
1 and 2 in a noisy environment. Specific noise model is characterized
through the index $k$. We have already assumed that Alice (Sender1)
and Bob (Sender2) wish to remotely prepare qubits $a_{1}|0\rangle+b_{1}\exp(i\phi_{1})|1\rangle$
and $a_{2}|0\rangle+b_{2}\exp(i\phi_{2})|1\rangle$ at the side of
Bob (Receiver1) and Alice (Receiver2), respectively. Thus, the expected
final state in the absence of noise is a product state where Alice (Receiver2)
will have qubit $a_{2}|0\rangle+b_{2}\exp(i\phi_{2})|1\rangle$ in
her possession and Bob (Receiver1) will have $a_{1}|0\rangle+b_{1}\exp(i\phi_{1})|1\rangle$
in his possession. As a consequence, in the ideal situation (i.e.,
in the absence of noise) in all successful cases of CBRSP the final state
would be \[
|T\rangle_{R_{1}R_{2}}=\left(a_{1}|0\rangle+b_{1}\exp(i\phi_{1})|1\rangle\right)\otimes\left(a_{2}|0\rangle+b_{2}\exp(i\phi_{2})|1\rangle\right).\]
For computational convenience, we assume that $a_{i}=\sin\theta_{i}$
and $b_{i}=\cos\theta_{i}$ with $i\in\{1,2\}$. Thus, \[
|T\rangle_{R_{1}R_{2}}=\text{sin}\theta_{1}\text{sin}\theta_{2}|00\rangle+\text{cos}\text{\ensuremath{\theta_{2}}}\text{sin}\ensuremath{\theta_{1}}\exp\left(i\phi_{2}\right)|01\rangle+\cos\text{\ensuremath{\theta_{1}}}\text{sin}\ensuremath{\theta_{2}}\exp\left(i\phi_{1}\right)|10\rangle+\text{cos}\text{\ensuremath{\theta_{1}}}\text{cos}\theta_{2}\exp\left(\phi_{1}+\phi_{2}\right)|11\rangle.\]
We can visualize the effect of noise by comparing the quantum
state $\rho_{k{\rm ,out}}$ prepared in the noisy environment with
the state $|T\rangle_{R_{1}R_{2}}$ using fidelity \begin{equation}
F=\langle T|\rho_{k,{\rm ou}t}|T\rangle,\label{eq:fidelity}\end{equation}
which is square of the usual definition of fidelity of two quantum
states $\rho$ and $\sigma$ defined as $F(\sigma,\rho)=Tr\sqrt{\sigma^{\frac{1}{2}}\rho\sigma^{\frac{1}{2}}}.$
In the present paper, we have used (\ref{eq:fidelity}) as the definition
of fidelity. 

\begin{figure}
\begin{centering}
\includegraphics[angle=0,scale=0.55]{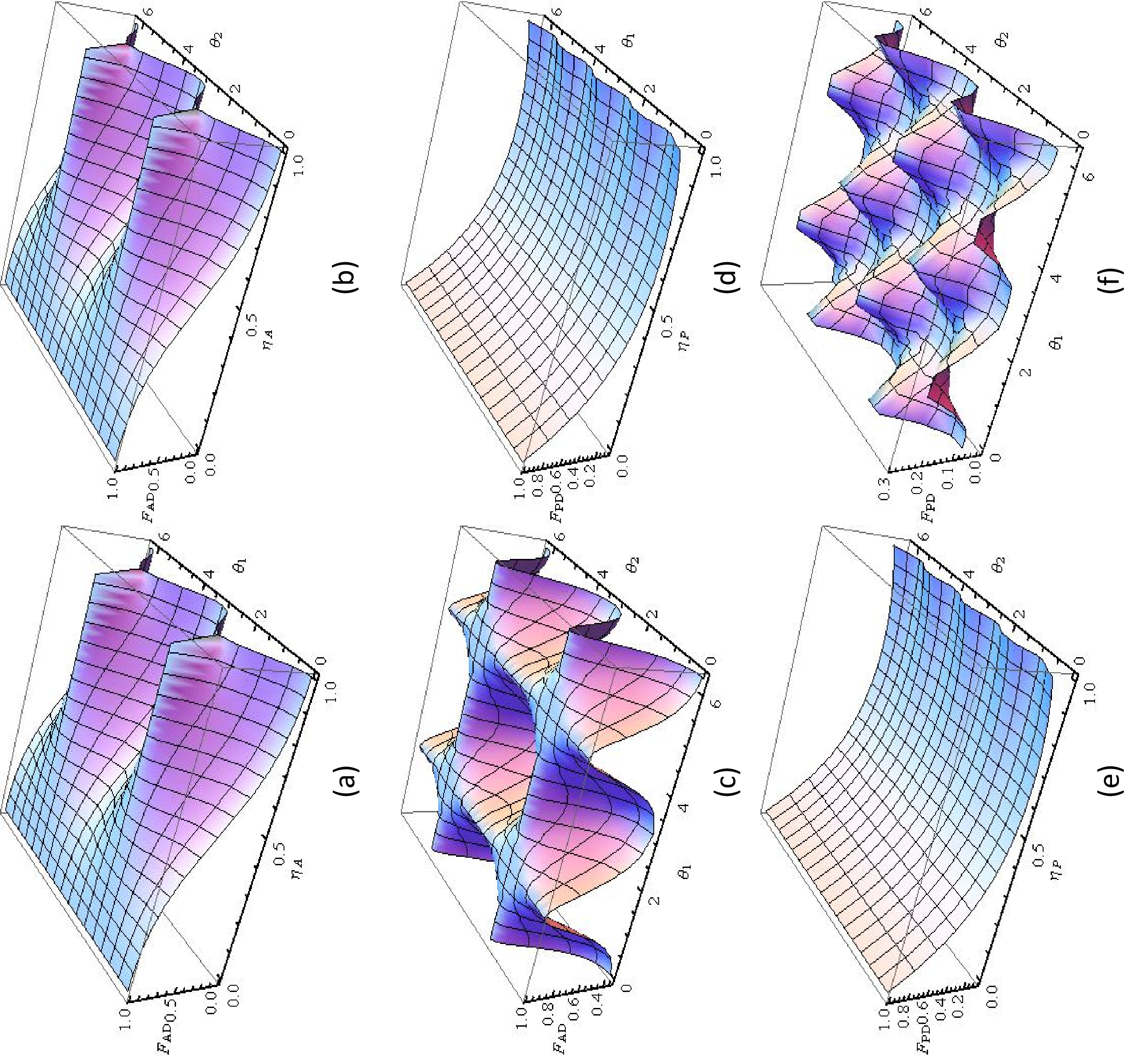}
\par\end{centering}
\caption{\label{fig:effect of noise}(Color online) Effect of noise on probabilistic
CBRSP scheme is visualized through variation of fidelity $F_{{\rm AD}}$
(for amplitude-damping noise model) and $F_{{\rm PD}}$ (for phase-damping
noise model) with respect to amplitude information of the states to
be prepared remotely (i.e., $\theta_{i}$) and decoherence rates (i.e.,
$\eta_{i}$) for various situations: (a) amplitude-damping noise with
$\theta_{2}=\frac{\pi}{4}$, (b) amplitude-damping noise with $\theta_{1}=\frac{\pi}{4}$
, (c) amplitude-damping noise with $\eta_{A}=0.5$, (d) phase-damping
noise with $\theta_{2}=\frac{\pi}{4}$, (e) phase-damping noise with
$\theta_{1}=\frac{\pi}{4}$, (f) phase-damping noise with $\eta_{P}=0.5$. }

\end{figure}
\begin{figure}

\begin{centering}
\includegraphics[scale=0.5]{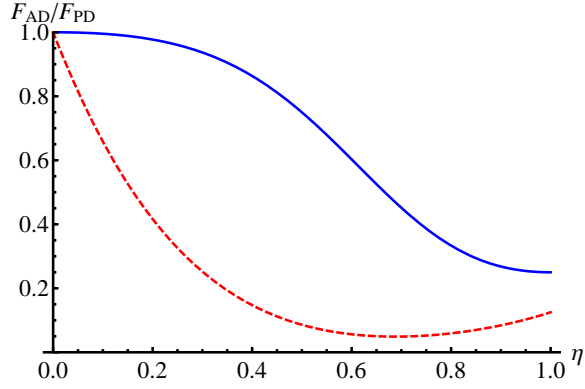}
\par\end{centering}

\caption{\label{fig:comparisno}(Color online) Comparison of the effect of amplitude-damping
noise (solid line) with phase damping noise (dashed line) by assuming
$\eta_{A}=\eta_{P}=\eta$ and $\theta_{1}=\theta_{2}=\frac{\pi}{4}$.
In this situation, fidelity for amplitude-damping noise is always
greater than that for the phase-damping noise for the same value of
decoherence rate $\eta.$ Fidelity for amplitude-damping noise monotonically
decreases with decoherence rate $\eta$, but fidelity for phase-damping
is found to increase with $\eta,$ after initially decreasing.}

\end{figure}

To study the effect of amplitude-damping and phase-damping noise models
we assume that the following specific quantum state of the general
from (\ref{eq:the 5-qubit state}) is used as a quantum channel for
probabilistic CBRSP

\begin{equation}
|\psi\rangle=\frac{1}{\sqrt{2}}\left(|\psi^{+}\rangle_{S_{1}R_{1}}|\psi^{+}\rangle_{S_{2}R_{2}}|0\rangle_{C_{1}}+|\phi^{-}\rangle_{S_{1}R_{1}}|\phi^{-}\rangle_{S_{2}R_{2}}|1\rangle_{C_{1}}\right).\label{eq:choosen qstate-for-PCBRSP}\end{equation}
For this particular choice of quantum channel, the above described
method of obtaining $\rho_{k,{\rm out}}$ yields

\begin{equation}
{\normalcolor }\begin{array}{l}
\rho_{A,{\rm out}}\\
=N_{A}\left(\begin{array}{cccc}
\rho_{A,11} & 4\text{sin}^{2}\text{\ensuremath{\theta_{1}}}\text{sin}2\text{\ensuremath{\theta_{2}}}\exp(\text{\ensuremath{-i\phi_{2}})} & 4\text{sin}2\text{\ensuremath{\theta_{1}}}\text{sin}^{2}\text{\ensuremath{\theta_{2}}}\exp(\text{\ensuremath{-i\phi_{1}})} & 2\text{sin}2\text{\ensuremath{\theta_{1}}}\text{sin}2\text{\ensuremath{\theta_{2}}}\exp(-i\phi_{12})\\
4\text{sin}^{2}\text{\ensuremath{\theta_{1}}}\text{sin}2\text{\ensuremath{\theta_{2}}}\exp(\text{\ensuremath{i\phi_{2}})} & 8\text{cos}^{2}\text{\ensuremath{\theta_{2}}}\text{sin}^{2}\text{\ensuremath{\theta_{1}}} & 2\text{sin}2\text{\ensuremath{\theta_{1}}}\text{sin}2\text{\ensuremath{\theta_{2}}}\exp(-i\Delta\phi) & 4\text{cos}^{2}\text{\ensuremath{\theta_{2}}}\text{sin}2\text{\ensuremath{\theta_{1}}}\exp(\text{\ensuremath{-i\phi_{1}})}\\
4\text{sin}2\text{\ensuremath{\theta_{1}}}\text{sin}^{2}\text{\ensuremath{\theta_{2}}}\exp(\text{\ensuremath{i\phi_{1}})} & 2\text{sin}2\text{\ensuremath{\theta_{1}}}\text{sin}2\text{\ensuremath{\theta_{2}}}{\normalcolor {\color{red}{\normalcolor \exp(i\Delta\phi)}}} & 8\text{cos}^{2}\text{\ensuremath{\theta_{1}}}\text{sin}^{2}\text{\ensuremath{\theta_{2}}} & 4\text{cos}^{2}\theta_{1}\text{sin}2\text{\ensuremath{\theta_{2}}}\exp(\text{\ensuremath{-i\phi_{2}})}\\
2\sin2\text{\ensuremath{\theta_{1}}}\text{sin}2\theta_{2}\exp(i\phi_{12}) & 4\text{cos}^{2}\text{\ensuremath{\theta_{2}}}\text{sin}2\text{\ensuremath{\theta_{1}}}\exp(\text{\ensuremath{i\phi_{1}})} & 4\text{cos}^{2}\ensuremath{\theta_{1}}\text{sin}2\text{\ensuremath{\theta_{2}}}\exp(\text{\ensuremath{i\phi_{2}})} & 8\text{cos}^{2}\text{\ensuremath{\theta_{1}}}\text{cos}^{2}\text{\ensuremath{\theta_{2}}}\end{array}\right),\end{array}\label{eq:rhoAout}\end{equation}
 and \begin{equation}
\begin{array}{lcl}
\rho_{P,{\rm out}} & = & \frac{(-1+\eta_{P})^{4}}{4}\\
 & \times & \left(\begin{array}{cccc}
4\text{sin}^{2}\text{\ensuremath{\theta_{1}}}\text{sin}^{2}\text{\ensuremath{\theta_{2}}} & 2\text{sin}^{2}\text{\ensuremath{\theta_{1}}}\text{sin}2\text{\ensuremath{\theta_{2}}}\exp(-i\phi_{2}) & 2\text{sin}2\text{\ensuremath{\theta_{1}}}\text{sin}^{2}\text{\ensuremath{\theta_{2}}}\exp(-i\phi_{1}) & \text{sin}2\text{\ensuremath{\theta_{1}}}\text{sin}2\text{\ensuremath{\theta_{2}}}\exp(-i\phi_{12})\\
2\text{sin}^{2}\text{\ensuremath{\theta_{1}}}\text{sin}2\text{\ensuremath{\theta_{2}}}\exp(i\phi_{2}) & \frac{4\left(1-2\eta_{P}+2\eta_{P}^{2}\right)^{2}}{(-1+\eta_{P})^{4}}\text{cos}^{2}\ensuremath{\theta_{2}}\text{sin}^{2}\text{\ensuremath{\theta_{1}}} & \text{sin}2\text{\ensuremath{\theta_{1}}}\text{sin}2\text{\ensuremath{\theta_{2}}}\exp(-i\Delta\phi) & 2\text{\ensuremath{\cos}}^{2}\text{\ensuremath{\theta_{2}}}\text{sin}2\text{\ensuremath{\theta_{1}}}\exp(-i\phi_{1})\\
2\text{sin}2\text{\ensuremath{\theta_{1}}}\text{sin}^{2}\text{\ensuremath{\theta_{2}}}\exp(i\phi_{1}) & \text{sin}2\text{\ensuremath{\theta_{1}}}\sin2\text{\ensuremath{\theta_{2}}}\exp(i\Delta\phi) & \frac{4\left(1-2\eta_{P}+2\eta_{P}^{2}\right)^{2}}{(-1+\eta_{P})^{4}}\text{cos}^{2}\text{\ensuremath{\theta_{1}}}\text{sin}^{2}\text{\ensuremath{\theta_{2}}} & 2\text{\ensuremath{\cos}}^{2}\text{\ensuremath{\theta_{1}}}\text{sin}2\text{\ensuremath{\theta_{2}}}\exp(-i\phi_{2})\\
\text{sin}2\text{\ensuremath{\theta_{1}}}\text{sin}2\text{\ensuremath{\theta_{2}}}\exp(i\phi_{12}) & 2\text{cos}^{2}\text{\ensuremath{\theta_{2}}}\sin2\text{\ensuremath{\theta_{1}}}\exp(i\phi_{1}) & 2\cos^{2}\text{\ensuremath{\theta_{1}}}\sin2\text{\ensuremath{\theta_{2}}}\exp(i\phi_{2}) & 4\text{cos}^{2}\text{\ensuremath{\theta_{1}}}\text{cos}^{2}\text{\ensuremath{\theta_{2}}}\end{array}\right)\end{array}\label{eq:rhoPout}\end{equation}
 where $\phi_{12}=\phi_{1}+\phi_{2},$ $\Delta\phi=(\phi_{1}-\phi_{2})$,
\[
\begin{array}{lcl}
\rho_{A,11} & = & \frac{1}{(-1+\eta)^{2}}\left(2-4\eta_{A}+6\eta_{A}^{2}+2\left(-1+2\eta_{A}+\eta_{A}^{2}\right)\text{cos}2\text{\ensuremath{\theta_{1}}}+\left(1-2\eta_{A}+3\eta_{A}^{2}\right)\text{cos}\left(2(\text{\ensuremath{\theta_{1}}}-\text{\ensuremath{\theta_{2}}})\right)-2\text{cos}2\text{\ensuremath{\theta_{2}}}+4\eta_{A}\text{cos}2\text{\ensuremath{\theta_{2}}}\right.\\
 & + & \left.2\eta_{A}^{2}\text{cos}2\text{\ensuremath{\theta_{2}}}+\text{cos}\left(2(\text{\ensuremath{\theta_{1}}}+\text{\ensuremath{\theta_{2}}})\right)-2\eta_{A}\text{cos}\left(2(\text{\ensuremath{\theta_{1}}}+\text{\ensuremath{\theta_{2}}})\right)+3\eta_{A}^{2}\text{cos}\left(2(\text{\ensuremath{\theta_{1}}}+\text{\ensuremath{\theta_{2}}})\right)\right),\end{array}\]
 and \[
N_{A}=\frac{(-1+\eta_{A})^{2}}{2\times\left(4-8\eta_{A}+6\eta_{A}^{2}+2\eta_{A}^{2}\text{cos}2\text{\ensuremath{\theta_{1}}}+\eta_{A}^{2}\text{cos}2(\text{\ensuremath{\theta_{1}}}-\text{\ensuremath{\theta_{2}}})+2\eta_{A}^{2}\text{cos}2\text{\ensuremath{\theta_{2}}}+\eta_{A}^{2}\text{cos}2(\text{\ensuremath{\theta_{1}}}+\text{\ensuremath{\theta_{2}}})\right)}.\]
Using (\ref{eq:fidelity}) and (\ref{eq:rhoAout}) we obtain the fidelity
of the quantum state prepared using the proposed probabilistic CBRSP
scheme under amplitude-damping noise as

\begin{equation}
F_{{\rm AD}}=\frac{64-128\eta_{A}+66\eta_{A}^{2}-2\eta_{A}^{2}\text{cos}4\text{\ensuremath{\theta_{1}}}+\eta_{A}^{2}\text{cos}\left(4(\text{\ensuremath{\theta_{1}}}-\text{\ensuremath{\theta_{2}}})\right)-2\eta_{A}^{2}\text{cos}4\text{\ensuremath{\theta_{2}}}+\eta_{A}^{2}\text{cos}\left(4(\text{\ensuremath{\theta_{1}}}+\text{\ensuremath{\theta_{2}}})\right)}{16\left(4-8\eta_{A}+6\eta_{A}^{2}+2\eta_{A}^{2}\text{cos}2\text{\ensuremath{\theta_{1}}}+\eta_{A}^{2}\text{cos}\left(2(\text{\ensuremath{\theta_{1}}}-\text{\ensuremath{\theta_{2}}})\right)+2\eta_{A}^{2}\text{cos}2\text{\ensuremath{\theta_{2}}}+\eta_{A}^{2}\text{cos}\left(2(\text{\ensuremath{\theta_{1}}}+\text{\ensuremath{\theta_{2}}})\right)\right)}.\label{eq:fidelity-Amp-damp-probab}\end{equation}
 Similarly, by using (\ref{eq:fidelity}) and (\ref{eq:rhoPout})
we obtain the fidelity of the quantum state prepared using the proposed
probabilistic CBRSP scheme under phase-damping noise as

\begin{equation}
\begin{array}{lcl}
F_{{\rm PD}} & = & \frac{1}{64}\left(64-256\eta_{P}+420\eta_{P}^{2}-328\eta_{P}^{3}+118\eta_{P}^{4}+6\eta_{P}^{2}\left(2-4\eta_{P}+3\eta_{P}^{2}\right)\text{cos}4\text{\ensuremath{\theta_{1}}}-16\eta_{P}^{2}\left(2-4\eta_{P}+3\eta_{P}^{2}\right)\text{cos}\left(2(\text{\ensuremath{\theta_{1}}}-\text{\ensuremath{\theta_{2}}})\right)\right.\\
 & + & 2\eta_{P}^{2}\text{cos}\left(4(\text{\ensuremath{\theta_{1}}}-\text{\ensuremath{\theta_{2}}})\right)-4\eta_{P}^{3}\text{cos}\left(4(\text{\ensuremath{\theta_{1}}}-\text{\ensuremath{\theta_{2}}})\right)+3\eta_{P}^{4}\text{cos}\left(4(\text{\ensuremath{\theta_{1}}}-\text{\ensuremath{\theta_{2}}})\right)+12\eta_{P}^{2}\text{cos}4\text{\ensuremath{\theta_{2}}}-24\eta_{P}^{3}\text{cos}4\text{\ensuremath{\theta_{2}}}\\
 & + & 18\eta_{P}^{4}\text{cos}4\text{\ensuremath{\theta_{2}}}-32\eta^{2}\text{cos}\left(2(\text{\ensuremath{\theta_{1}}}+\text{\ensuremath{\theta_{2}}})\right)+64\eta_{P}^{3}\text{cos}\left(2(\text{\ensuremath{\theta_{1}}}+\text{\ensuremath{\theta_{2}}})\right)-48\eta_{P}^{4}\text{cos}\left(2(\text{\ensuremath{\theta_{1}}}+\text{\ensuremath{\theta_{2}}})\right)\\
 & + & \left.2\eta_{P}^{2}\text{cos}\left(4(\text{\ensuremath{\theta_{1}}}+\text{\ensuremath{\theta_{2}}})\right)-4\eta_{P}^{3}\text{cos}\left(4(\text{\ensuremath{\theta_{1}}}+\text{\ensuremath{\theta_{2}}})\right)+3\eta_{P}^{4}\text{cos}\left(4(\text{\ensuremath{\theta_{1}}}+\text{\ensuremath{\theta_{2}}})\right)\right).\end{array}\label{eq:fidelity-phase-damp-probab}\end{equation}
From (\ref{eq:fidelity-Amp-damp-probab}) and (\ref{eq:fidelity-phase-damp-probab})
we can see that both the fidelities $F_{{\rm AD}}$ for amplitude-damping
noise and $F_{{\rm PD}}$ for phase-damping noise depend only on the
decoherence rate $\eta_{k}$ and the amplitude information (i.e.,
$a_{i}$ and $b_{i}$) of the states that were attempted to be prepared
remotely and fidelities are independent of the corresponding phase
information $\phi_{i}.$ A similar observation was recently reported
in Ref. \cite{RSP-with-noise} in the context of JRSP in noisy environments. The method adopted to study the effect of
noise in the present paper and in Ref. \cite{RSP-with-noise} is quite
general and can be easily applied to other schemes of quantum communication
in general and to the schemes of bidirectional quantum communication
in particular. For example, if we wish to extend the present discussion
to the case of the deterministic CBRSP scheme described in Section \ref{sec:Deterministic-controlled-bidirectional}
then we have to use a 7-qubit state, but the Kraus operators would
still operate on the same qubits as in the case of the probabilistic
CBRSP as the additional qubits used for deterministic CBRSP are prepared
locally by the senders and these qubits are not exposed to the noisy
environment. However, in case we wish to implement the scheme of deterministic
CJBRSP as described in Section \ref{sec:Deterministic-controlled-joint},
we have to apply Kraus operators on 6 qubits (except the qubit of
controller) of a 7-qubit quantum channel of the general form (\ref{eq:7-qubit-3}).
Fidelities for deterministic CBRSP, CJBRSP and/or controlled bidirectional
teleportation can be obtained easily by following the procedure adopted
in this work. However, here we restrict ourselves to the case of probabilistic
CBRSP alone. In case the probabilistic CBRSP scheme, realized using
the quantum channel (\ref{eq:choosen qstate-for-PCBRSP}), is exposed
to different noisy environments, then the fidelity corresponding to
various noise models would depend on the corresponding parameters, as shown
in Fig. \ref{fig:effect of noise}. Specifically, Fig. \ref{fig:effect of noise}
a-c (d-f) clearly illustrates the effect of amplitude-damping (phase-damping)
noise on the fidelity $F_{{\rm AD}}$ ($F_{{\rm PD}})$ and variation
of the fidelity with $\theta_{i}$ (or equivalently $a_{i}$ and $b_{i}$
) and decoherence rate $\eta_{k}.$ We can easily observe that fidelity
$F_{{\rm AD}}$ always decreases with decoherence $\eta_{A},$ (c.f.
Fig. \ref{fig:effect of noise} a-b), but similar character is not
observed in phase-damping channel where we can observe that initially
fidelity decreases with $\eta_{P}$ and after a point it starts increasing
(c.f. Fig. \ref{fig:effect of noise} d-e). These characteristics
can be visualized more clearly in Fig. \ref{fig:comparisno}, where
we have compared the effect of amplitude-damping noise with phase
damping noise by assuming $\eta_{A}=\eta_{P}=\eta$ and $\theta_{1}=\theta_{2}=\frac{\pi}{4}$.
In this situation, fidelity of amplitude-damping channel (solid line
in Fig. \ref{fig:comparisno}) is always more than that of the phase-damping
channel (dashed line in Fig. \ref{fig:comparisno}) for the same value
of decoherence rate $\eta.$ Thus we see that information loss is less when the travel qubits are
transferred through the amplitude-damping channel as compared to the phase-damping channel, in
consistence with the work in \cite{RSP-with-noise} where similar considerations were applied to a JRSP process.

\section{Conclusion \label{sec:Conclusion}}

We have provided protocols of probabilistic CBRSP, deterministic CBRSP
and deterministic CJBRSP. Interestingly, the probabilistic CBRSP requires a
lesser amount of classical communications compared to the BCST schemes
\cite{bi-directional-ourpaper} which are deterministic. This advantage
of lesser classical communication is lost in the deterministic CBRSP.
This observation is analogous to the one-directional case. However,
the operations used in deterministic CBRSP are such that it can be
modified quickly into the protocol of deterministic CJBRSP. Interestingly,
it is shown that the above protocols can be realized using infinitely
many alternative quantum channels. Further, the deterministic protocols
described above can also be turned in to probabilistic protocols of
bidirectional RSP by considering the $GHZ$ states shared by the sender(s)
and receiver to be non-maximally entangled. In such a situation we
will obtain a controlled bidirectional version of the recently proposed
RSP scheme of Wei et al. \cite{probabilistic-RSP-with-non-maximal-channel}.
The protocols presented here are also interesting because of its potential
applications in several practical situations discussed in earlier
works on RSP. Further, the presented protocol is experimentally realizable
using presently available technologies and for the first time, to the best of our knowledge, effect
of noise on a bidirectional quantum communication protocol is described. 
The effect of amplitude-damping  and phase-damping
noise, on our protocols, makes the present study much more realistic compared to the existing works 
as in practice we cannot have a noise-free quantum
channel. In addition, the method used here for the study of effect
of noise is quite general and it is possible to apply this approach
to study the effect of noise in other similar schemes of quantum communication.

\textbf{Acknowledgment}: AP thanks Department of Science and Technology
(DST), India for support provided through the DST project No. SR/S2/LOP-0012/2010. He also thanks N. B. An for some technical discussion.


\begin{thebibliography}{41}
\bibitem{Bennett}Bennett, C. H., et al.: Phys. Rev. Lett. \textbf{70},
1895 (l993)

\bibitem{my book}Pathak, A.: Elements of quantum computation and
quantum communication. CRC Press, Boca Raton, USA (2013)

\bibitem{Hillery}Hillery, M., Buzek, V., Bertaiume, A.: Phys. Rev.
A \textbf{59}, 1829 (1999)

\bibitem{Ct}Karlsson, A., Bourennane, M.: Phys. Rev. A \textbf{58},
4394 (1998)

\bibitem{A.Pathak}Pathak, A., Banerjee, A.: Int. J. Quantum Info.
\textbf{9}, 389 (2011)

\bibitem{hierarchical}Wang, X. W.\emph{, }et al.: Opt. Commun. \textbf{283},
1196 (2010)

\bibitem{Shukla}Shukla, C., Pathak, A.: Phys. Lett. A \textbf{377},
1337 (2013)

\bibitem{Pati-original}Pati, A. K.: Phys. Rev. A \textbf{63}, 014302
(2000)

\bibitem{LO-RSP}Lo, H. K.: Phys. Rev. A \textbf{62}, 012313 (2000)

\bibitem{Bennet-RSP}Bennett, C. H., DiVincenzo, D. P., Shor, P. W.,
Smolin, J. A., Terhal, B. M., Wootters, W. K.: Phys. Rev. Lett. \textbf{87},
077902 (2001)

\bibitem{Zha}Zha, X.-W, Zou, Z.-C., Qi, J.-X., Song, H.-Y.: Int.
J. Theor. Phys. \textbf{52}, 1740 (2013)

\bibitem{Zha II}\textcolor{black}{Zha, X.-W., Song, H.-Y., Ma, G.-L.:
quant-ph/1006.0052, (2010)}

\bibitem{bi-directional-ourpaper}Shukla, C., Banerjee, A., Pathak,
A.: Int. J. Theor. Phys.\textbf{ 52}, 3790 (2013)

\bibitem{7qubit}Duan, Y.-J., Zha, X.-W., Sun, X.-M., Xia, J.-F.:
Int. J. Theor. Phys. \textbf{53}, 2697 (2014)

\bibitem{sixqubit1}Duan, Y.-J., Zha, X.-W.: Int. J. Theor. Phys.
DOI 10.1007/s10773-014-2131-8 (2014)

\bibitem{six-qubit-3}Chen, Y.: Int. J. Theor. Phys. DOI 10.1007/s10773-014-2221-7
(2014)

\bibitem{sixqubiit2}Fu, H.-Z., Tian, X.-L., Hu, Y.: Int. J. Theor.
Phys. \textbf{53}, 1840 (2014)

\bibitem{six-qubit-4}An, Y.: Int. J. Theor. Phys. \textbf{52}, 3870
(2013)

\bibitem{Li}Li, Y.-h., Nie, L.-p.: Int. J. Theor. Phys. \textbf{52}
1630 (2013)

\bibitem{5-qubit-c-qsdc}Li, Y.-h., Li, X.-l., Sang, M.-h., Nie, Y.-y.,
Wang, Z.-s.: Quantum Inf. Process. \textbf{12}, 3835 (2013)

\bibitem{CRSP-hwang}Liu, L. L., Hwang, T.: Quantum Inf. Process.
\textbf{13}, 1639 (2014)

\bibitem{JRSP-cluster-like}Wang, D., Ye, L.: Int. J. Theor. Phys.\textbf{
52}, 3075 (2013)

\bibitem{JRSP with paasive receiver}Chen, Q. Q., Xia, Y., An, N.
B.: Phys. Scr. \textbf{87}, 025005 (2013)

\bibitem{j-rsp-multi}Wang, D., Ye, L.: Quantum Inf. Process. \textbf{12},
3223 (2013)

\bibitem{J. A. Vaccaro}Huelga, S. F. et al., Phys. Rev. A \textbf{63},
042303 (2001)

\bibitem{J. A. Vaccaro-1}Huelga, S. F., Plenio, M. B., Vaccaro, J.
A.: Phys. Rev. A\textbf{ 65}, 042316 (2002)

\bibitem{ba-an-deterministic}An, N. B., Cao, T. B., Nung, V. D.,
Kim, J.: Advances in Natural Sciences: Nanoscience and Nanotechnology
\textbf{2}, 035009 (2011)

\bibitem{RSP-1-GHZ}Dai, H. Y., Chen, P. X., Liang, L. M., Li, C.
Z.: Phys. Lett. A \textbf{355}, 285 (2006)

\bibitem{RSP-multiparitite-ghz}Ma, P.-C., Zhan, Y.-B.: Chin. Phys.
B \textbf{17}, 445 (2008)

\bibitem{RSP-4-qubit-cluster-type}Ma, P.-C., Zhan, Y.-B.: Opt. Commun.
\textbf{283}, 2640 (2010)

\bibitem{RSP-cluster-type}Zhan, Y.-B., Fu, H., Li, X.-W., Ma, P.-C.:
Int. J. Theor. Phys. \textbf{52}, 2615 (2013)

\bibitem{RSP-arbitrary-bipartite-usingGHZ-type}Peng, J.-Y., Luo,
M.-X., Mo, Z.-W.: Quantum Inf. Process. \textbf{12}, 2325 (2013)

\bibitem{JRSP-W}Chen, Q. Q., Xia, Y., Song, J., An, N. B.: Phys.
Lett. A \textbf{374}, 4483 (2010)

\bibitem{JRSP-W-state}An, N. B.: Opt. Commun. \textbf{283}, 4113
(2010)

\bibitem{Ba An -RSP}An, N. B., Kim, J.: J. Phys. B. \textbf{41},
095501 (2008)

\bibitem{crsp1}Wang, Z. Y., Liu, Y. M., Zuo, X. Q., Zhang, Z. J.:
Theor. Phys. (Beijing, China) \textbf{52}, 235 (2009)

\bibitem{cjrsp-1}Guan, X. W., Chen, X. B., Yang, Y. X.: Int. J. Theor.
Phys.\textbf{ 51}, 3575 (2012)

\bibitem{JRSP-BA-An}An, N. B., Cao, T. B., Nung, V. D.: Phys. Lett.
A \textbf{375}, 3570 (2011)

\bibitem{ba-An-remote-state}Cao, T. B., An, N. B.: Advances in Natural
Sciences: Nanoscience and Nanotechnology \textbf{5}, 015003 (2014)

\bibitem{turchette}  Turchette, Q. A., {\it et al.}: Phys. Rev. A \textbf{62},
053807 (2000); C. J. Myatt, {\it et al.}: Nature \textbf{403}, 269 (2000)

\bibitem{srikgp} Banerjee, S.,  Srikanth, R.:  Eur. Phys. J. D \textbf{46},
335 (2008); Srikanth, R., and  Banerjee, S.,: Phys. Rev. A \textbf{77},
155420 (2008)

\bibitem{ghoshqnd}  Banerjee, S.,  Ghosh, R.: J. Phys. A: Math. Theo. \textbf{40},
13735 (2007)

\bibitem{RSP-with-noise}Guan, X.-W., Chen, X.-B., Wang, L.-C., Yang,
Y.-X.: Int. J. Theor. Phys. \textbf{53}, 2236 (2014)

\bibitem{probabilistic-RSP-with-non-maximal-channel}Wei, J., Dai,
H.-Y., Zhang, M.: Quantum Inf. Process. DOI 10.1007/s11128-014-0799-6
(2014) 
\end{thebibliography}
\end{document}